\documentclass[11pt]{article}
\usepackage{moriond,epsfig}
\usepackage{overpic}%more figures in the same floating object

\def\be{\begin{equation}}
\def\ee{\end{equation}}
\def\bea{\begin{eqnarray}}
\def\eea{\end{eqnarray}}

\newcommand{\ttbar}{\ensuremath{t\bar{t}}}

\newcommand{\etmissx}{\ensuremath{E \kern-0.6em\slash_{\rm x}}}
\newcommand{\etmissy}{\ensuremath{E \kern-0.6em\slash_{\rm y}}}

\newcommand{\gtop}{\ensuremath{\Gamma_t}}
\newcommand{\gtwb}{\ensuremath{\Gamma_{t\rightarrow Wb}}}
\newcommand{\btwb}{\ensuremath{\mathcal B_{t\rightarrow Wb}}}
\newcommand{\sigt}{\ensuremath{\sigma_{t-{\rm channel}}}}
\newcommand{\afb}{\ensuremath{A_{\rm fb}}}

\newcommand{\GeV}{\ensuremath{\textnormal{GeV}}}
\newcommand{\TeV}{\ensuremath{\textnormal{TeV}}}

\newcommand{\dif}{\ensuremath{{\rm d}}}

\newcommand{\fb}{\ensuremath{{\rm fb}^{-1}}}

\newcommand{\etal}{{\em et. al.}}

\begin{document}
\vspace*{4cm}
\title{MEASUREMENTS OF THE PROPERTIES OF THE TOP QUARK}

\author{{\sc O. Brandt} on behalf of the {\sc CDF} and {\sc D0 Collaborations}}

\address{II. Physikalisches Institut, Friedrich-Hund-Platz 1,\\
G\"ottingen, Germany}

\maketitle\abstracts{
We review recent measurements of the properties of the top quark: the decay width of the top quark, of spin correlations between the top and the antitop quarks in $\ttbar$ production, the $W$ boson helicity in top decays, the strong colour flow in $\ttbar$ events, and the asymmetry of $\ttbar$ production due to the strong colour charge. The measurements are performed on data samples of up to 5.4~\fb\ of integrated luminosity acquired by the CDF and D0 collaborations in Run II of the Fermilab Tevatron $p\bar p$ collider at a centre-of-mass energy of $\sqrt s=1.96~\TeV$. 
%In particular, the measurements of the decay width of the top quark, of the spin correlations between the top  will be presented,
\vspace{2mm}\\
PACS {\tt 14.65.Ha} -- Top quarks.
}

\section{Introduction}
The pair-production of the top quark was discovered in 1995 by the CDF and D0 experiments~\cite{bib:topdiscovery} at the Fermilab Tevatron proton-antiproton collider. Observation of the electroweak production of single top quarks was presented only two years ago~\cite{bib:singletop}. The large top quark mass~\cite{bib:mass} and the resulting Yukawa coupling of about $0.996\pm0.006$ indicates that the top quark could play a crucial role in electroweak symmetry breaking. Precise measurements of the properties of the top quark provide a crucial test of the consistency of the standard model (SM) and could hint at physics beyond the SM. Only a small fraction of those measurements will be presented in the following, while their full listing can be found in~Refs.~\cite{bib:toprescdf,bib:topresd0}.

At the Tevatron, top quarks are mostly produced in pairs via the strong interaction. %, in about 85\% of the cases via $q\bar q'$ annihilation and in about 15\% via gluon-gluon fusion. 
At the time of the conference, about 9.5 fb$^{-1}$ of integrated luminosity per experiment were recorded by CDF and D\O, which corresponds to about 70k produced $\ttbar$ pairs. In the framework of the SM, the top quark decays to a $W$ boson and a $b$ quark nearly 100\% of the time, resulting in a $W^+W^-b\bar b$ final state from top quark pair production. 
%One of the challenges in measuring the top quark mass is the assignment of reconstruced leptons, jets, and missing transverse energy $\etmiss$ to partons, which, in absence of jet charge and flavour identification, can lead to several possible combinations. 
Thus, $\ttbar$ events are classified according to the $W$ boson decay channels as ``dileptonic'', ``all--jets'', or ``lepton+jets''. More details on the channels and their experimental challenges can be found in Ref.~\cite{bib:xsec}, while the electroweak production of single top quarks is reviewed in Ref.~\cite{bib:singletoptalk}.

\section{Measurement of the decay width of the top quark}
The D0 collaboration extracted the total decay width of the top quark~\cite{bib:width_d0}, $\gtop=\gtwb/\btwb$, from the partial decay width $\gtwb$ measured using the $t$-channel cross section for single top quark production, and from the branching fraction $\btwb$ measured in $\ttbar$ events using up to 2.3~\fb\ of integrated luminosity. This extraction is made under the assumption that the electroweak coupling in top quark production is identical to that in the decay. In this spirit, only the $t$-channel single top-quark production cross-section was used, since contributions from physics beyond the SM are expected to have different effects on $t$- and $s$-channels. Another theoretical input is the validity of next-to-leading order~(NLO) calculations of $\gtwb^{\rm SM}$ and $\sigt^{\rm SM}$, which enter the calculation as follows:
$
\gtop=\frac{\sigt}{\btwb}\times\frac{\gtwb^{\rm SM}}{\sigt^{\rm SM}}.
$
Properly taking into account all systematic uncertainties and their correlations among the measurements of \gtwb\ and \sigt, D0 finds $\gtop=1.99^{+0.69}_{-0.55}~\GeV$, which translates into a top-quark lifetime of $\tau_t=(3.3^{+1.3}_{-0.9})\times10^{-25}$~s, in agreement with the SM expectation. This constitutes the world's most precise indirect determination of \gtop\ to date. CDF has performed a direct measurement of $\gtop$, and set a limit $\gtop<7.6~\GeV$ at 95\% confidence level~(CL)~\cite{bib:width_cdf}.

\section{Measurement of spin correlations between top and antitop quarks}
\begin{figure}
\centering
\begin{overpic}[height=0.3\textwidth]{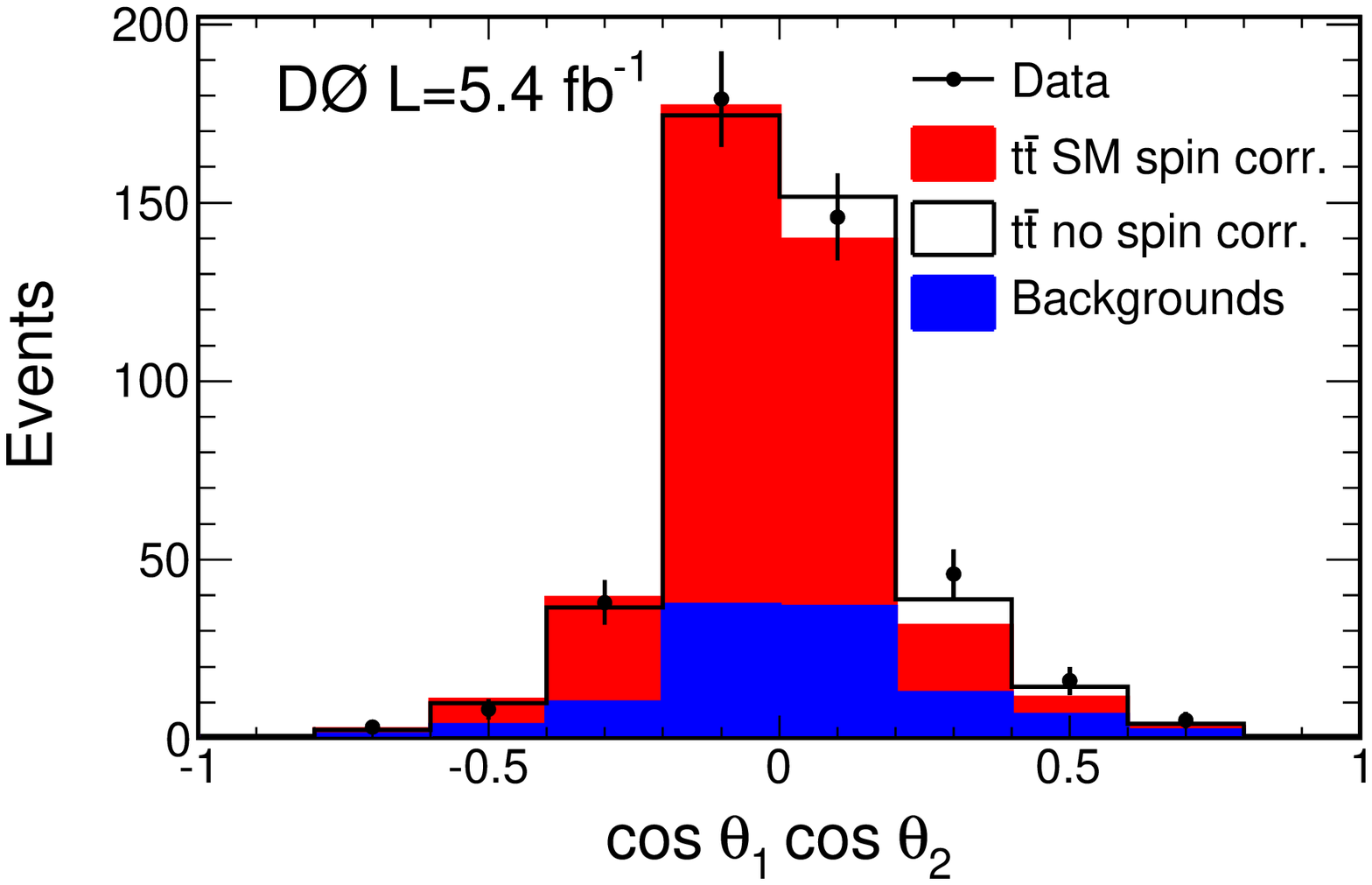}
\put(-1,2){(a)}
\end{overpic}
\qquad
\begin{overpic}[height=0.3\textwidth]{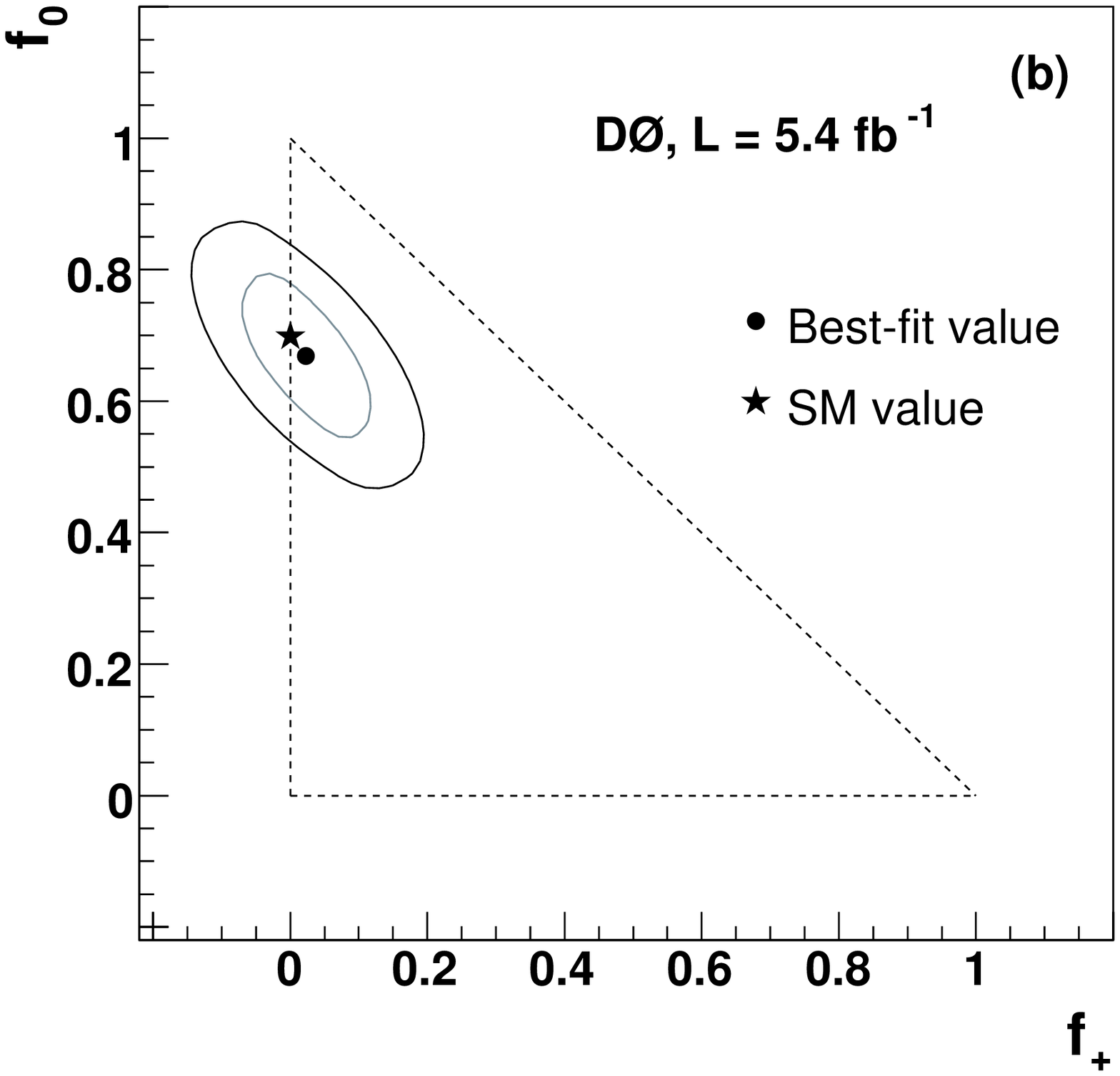}
\put(-1,2){(b)}
\end{overpic}
\caption{
\label{fig:spin}
{\bf(a)} The distribution in $\cos\theta_1\cos\theta_2$ for the entire dilepton sample with a dataset of 5.4 \fb\ at D0. The summed \ttbar\ signal with (without) NLO QCD spin correlations is shown as a red (open) histogram. {\bf(b)} The result of the model-independent $W$ boson helicity fit with a dataset of 5.4 \fb\ at D0. The ellipses indicate the 68\% and 95\% CL contours.
}
\end{figure}

While the top quarks are unpolarised in \ttbar\ production at hadron colliders, the orientation of their spins is correlated. In contrast to other quarks, this correlation is not affected by fragmentation due to the short life time of the top quark, and is thus reflected in its decay products. The spin correlation is defined as
$
C\equiv\frac{N_{\uparrow\uparrow}+N_{\downarrow\downarrow}-N_{\uparrow\downarrow}-N_{\downarrow\uparrow}}{N_{\uparrow\uparrow}+N_{\downarrow\downarrow}+N_{\uparrow\downarrow}+N_{\downarrow\uparrow}}
$
with $-1<C<+1$, and depends on the choice of the spin quantisation axis, which, for the measurements presented here, is defined by the direction of the incoming proton beam (``beamline axis''). 
%At threshold, large spin correlations are expected if the incoming partons combine into a $^3$S$_1$ state ($q\bar q$ production, 85\% at the Tevatron), while no spin correlations are expected for the $^1$S$_0$ state ($gg$ production, 15\%), which is strikingly different from the situation at the LHC, where the balance between the two is inverted.
D0 performed a measurement of $C$ in the dilepton channel using 5.4~\fb\ of data~\cite{bib:spin_d0} by analysing the distribution $1/{\sigma_{\ttbar}}\times{\dif^2\sigma_{\ttbar}}/{\dif\cos\theta_1\dif\cos\theta_2}=0.25\cdot(1-C\cos\theta_1\cos\theta_2)\,,
$
%\[
%\frac1{\sigma_{\ttbar}}\frac{\dif^2\sigma_{\ttbar}}{\dif\cos\theta_1\dif\cos\theta_2}=\frac14(1-C\cos\theta_1\cos\theta_2)\,,
%\]
shown in Fig.~\ref{fig:spin}~(a), and found $C=0.10\pm0.45$. Here, $\theta_{1,2}$ are the angles between the three-momenta of $\ell^+$ (resp. $\ell^-$) in the $t$ (resp. $\bar t$) rest frames and the quantisation axis. CDF performed a similar measurement in the lepton$+$jets final states using 5.3~\fb\ of data~\cite{bib:spin_cdf}, and found $C=0.72\pm0.69$. Both measurements are in agreement with the NLO QCD prediction of $C=0.78^{+0.03}_{-0.04}$. The above analyses are complementary to the LHC, where the $gg$ production mechanism dominates, and a much smaller $C$ is expected.

\section{Measurement of $W$ boson helicity in $\mathbf{t\bar t}$ events}
In the SM, the top quark decays into a $W$ boson and a $b$ quark with a probability of $>99.8\%$, where the on-shell $W$ boson can be in a left-handed, longitudinal, and right-handed helicity state. A NLO calculation within the SM of the corresponding helicity fractions predicts $f_-=0.301,\,f_0=0.698,$ and $f_+=4.1\times10^{-4}$, respectively. A significant deviation from the SM expectation would indicate a contribution from new physics. D0 has measured the $f_0$ and $f_+$ helicity fractions in dilepton and lepton$+$jets final states using up 5.4~\fb\ of data~\cite{bib:whel_d0}. A model-independent fit to the distribution in $\cos\theta^*$, where $\theta^*$ is the angle between the three-momentum of the top quark and the down-type fermion, yields $f_0=0.669\pm0.102$ and $f_+=0.023\pm0.053$ as shown in Fig.~\ref{fig:spin}~(b), in agreement with the SM expectation. CDF performed a similar model-independent measurement in dilepton events using 4.8~\fb\ of data, and found $f_0=0.78\pm0.21$ and $f_+=-0.12\pm0.12$~\cite{bib:whel_cdf}.

\section{Measurement of the strong colour flow in $\mathbf{t\bar t}$ events}
At leading order in $\alpha_s$, the strong colour charge can be traced from final to initial state partons, i.e. final state partons originating from the same initial state parton are ``colour-connected''. The potential energy of this colour-connection string is released in form of hadroproduction, and can be detected in the calorimeter. This can serve to separate processes which otherwise appear similar, like e.g. the decay of the Higgs boson and $g\rightarrow b\bar b$ production, which correspond~to a colour singlet and octet, respectively. D0 performed the first measurement of the colour representation of the hadronically decaying $W$~bosons in $\ttbar$ events in the lepton$+$jets channel~\cite{bib:colour}. Using calorimeter-based topological observables and a dataset corresponding to 5.3~\fb, D0~finds a fraction of $W$~bosons in colour-singlet configuration of $0.56\pm0.42$, in agreement with the~SM.

%\section{Measurement of the charge of the top quark}
%\cite{bib:charge}

\section{Measurement of the strong colour charge production asymmetry in $\mathbf{t\bar t}$ events}
\begin{figure}
\centering
\begin{overpic}[clip,height=3.7cm]{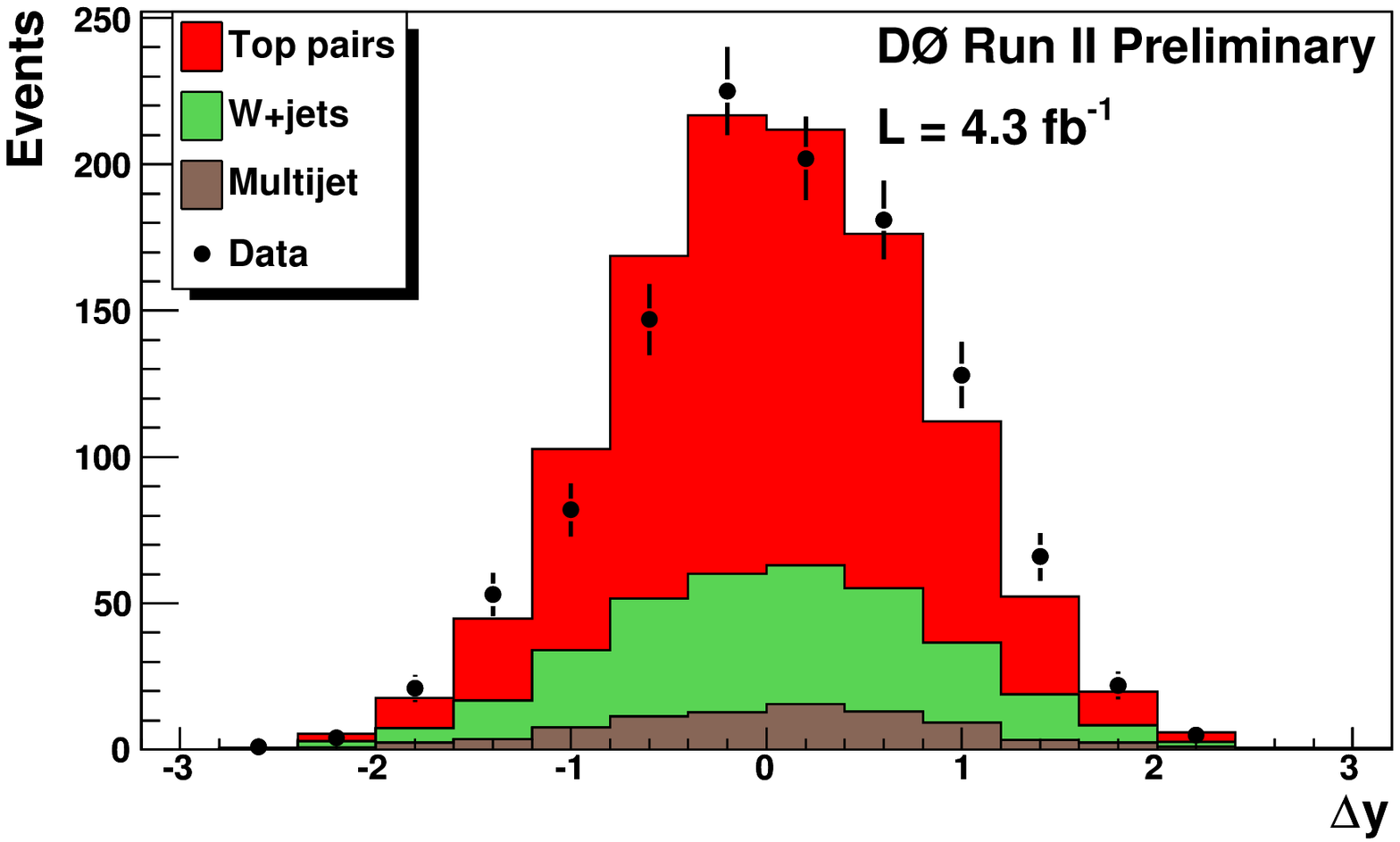}
\put(-1,2){(a)}
\end{overpic}
\quad
\begin{overpic}[clip,height=3.7cm]{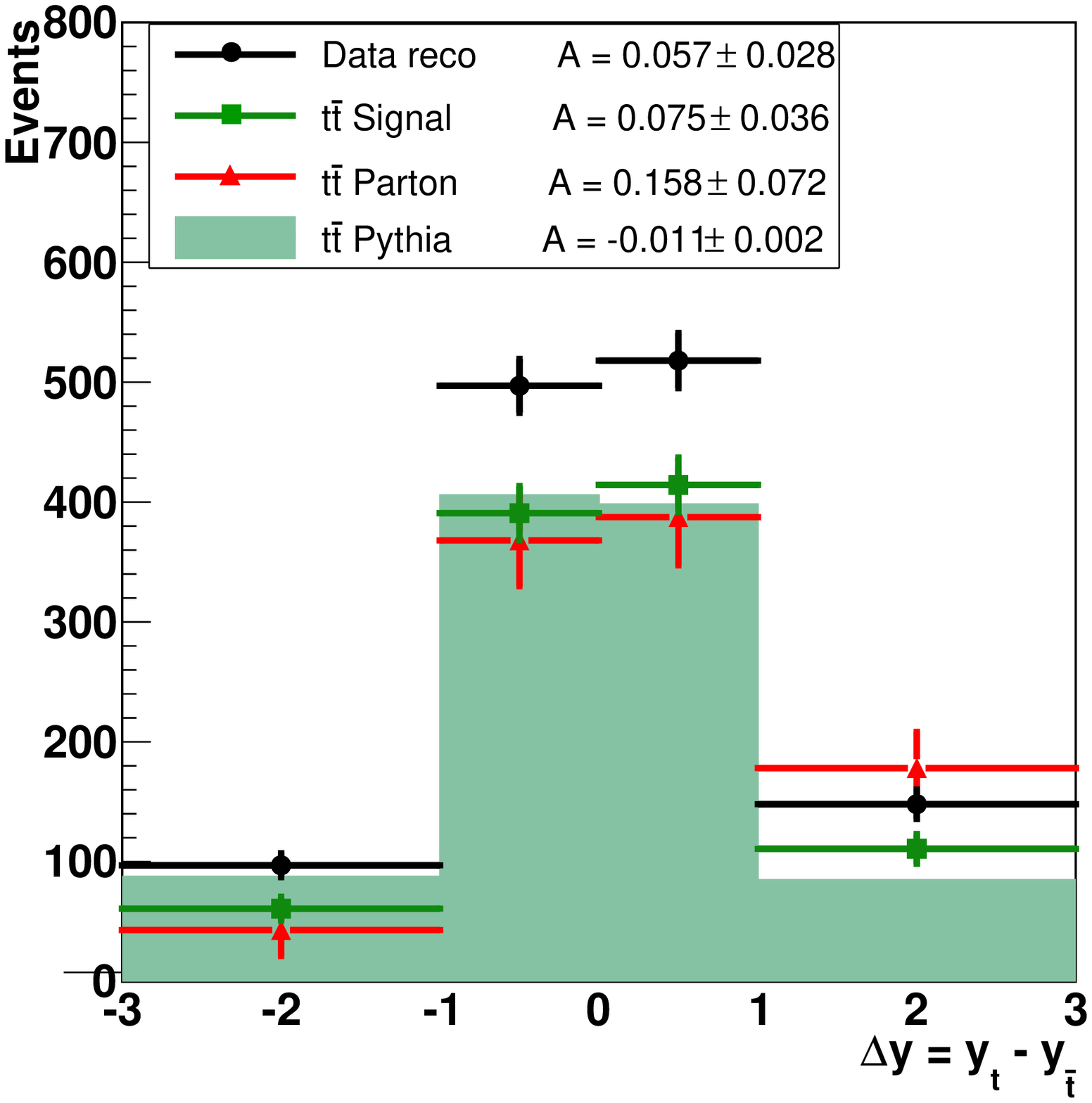}
\put(-1,2){(b)}
\end{overpic}
\quad
\begin{overpic}[clip,height=3.7cm]{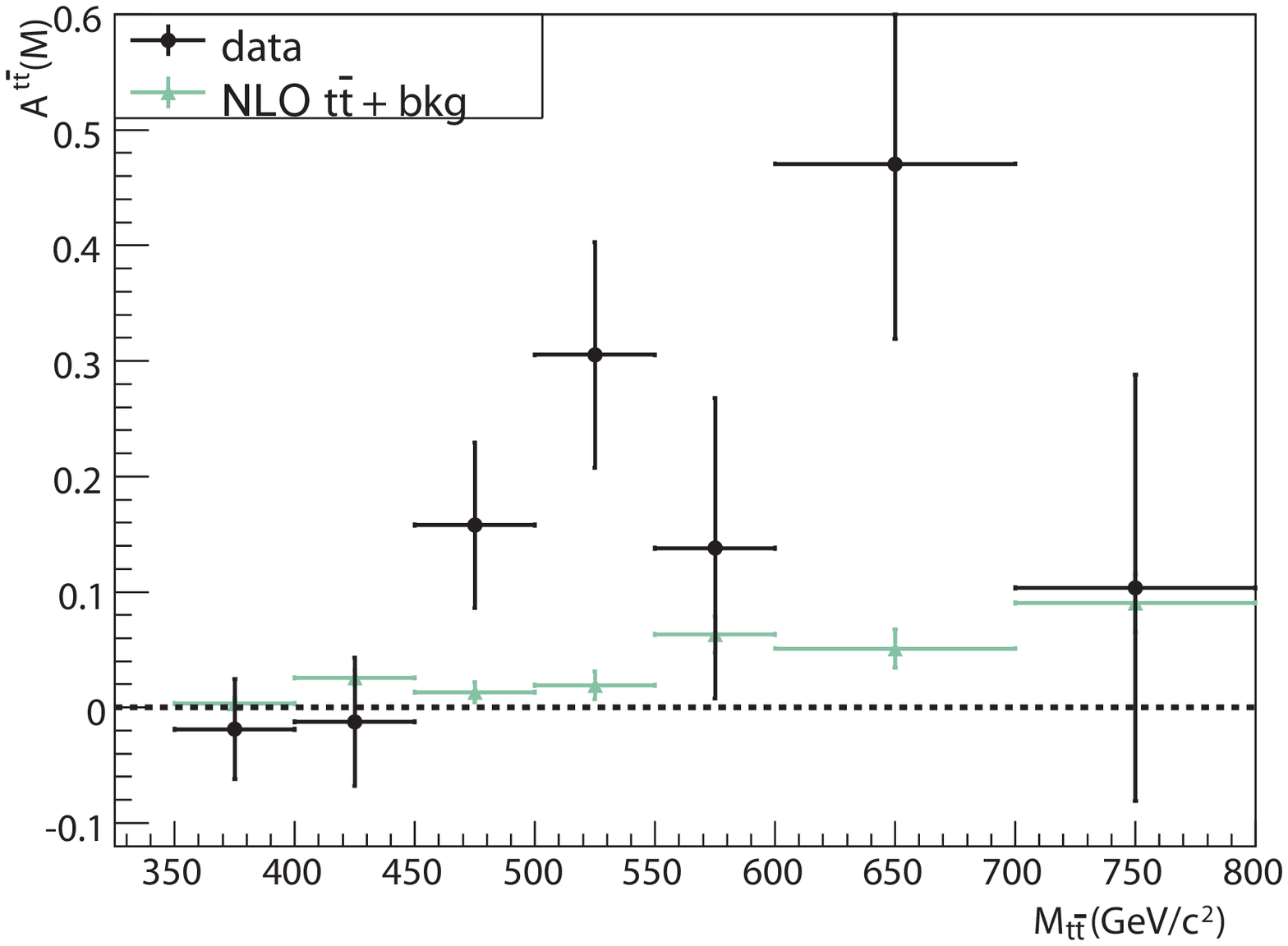}
\put(-1,2){(c)}
\end{overpic}
%\qquad
%\begin{overpic}[clip,height=0.3\textwidth]{fig/asymm_summ_cdf}
%\put(-1,2){(d)}
%\end{overpic}
\caption{\label{fig:asymm}
{\bf(a)} The distribution in $\Delta y$ for data and signal plus background fit to the discriminant with a dataset of 4.3 \fb\ at D0.
{\bf(b)} The distribution in $\Delta y$ for all correction levels with a dataset of 5.3 \fb\ at CDF. The corresponding $\afb$ values are summarised in the legend.
{\bf(c)} The distribution in $\afb$ in the \ttbar\ rest frame versus $M_{\ttbar}$ is shown for 5.3 \fb\ of CDF data and {\sc mc\@nlo} \ttbar\ signal plus background.
%{\bf(d)} $\afb$ after corrections to parton level, as measured by CDF with a dataset of 5.3 \fb, for events with $M_{\ttbar}$ below and above $450~\GeV$.
}
\end{figure}
%
%\begin{figure}
%\centering
%\begin{overpic}[clip,height=0.3\textwidth]{fig/asymm_vs_mtt_cdf}
%\put(-1,2){(a)}
%\end{overpic}
%\qquad
%\begin{overpic}[clip,height=0.3\textwidth]{fig/asymm_summ_cdf}
%\put(-1,2){(b)}
%\end{overpic}
%\caption{\label{fig:asymm_summ}
%{\bf(a)} The distribution in $\afb$ in the \ttbar\ rest frame versus $M_{\ttbar}$ is shown for 5.3 \fb\ of CDF data and {\sc mc\@nlo} \ttbar\ signal plus background.
%{\bf(b)} $\afb$ after corrections to parton level, as measured by CDF with a dataset of 5.3 \fb, for events with $M_{\ttbar}$ below and above $450~\GeV$.
%}
%\end{figure}
%
In the SM, the pair production of top quarks in $p\bar p$ collisions at LO is symmetric under charge conjugation. NLO calculations %within the SM 
predict a small forward-backward asymmetry $\afb$ of the order of 5\% in the $\ttbar$ rest frame, which is due to a negative contribution from the interference of diagrams for initial and final state radiation, and a (larger) positive contribution from the interference of box and tree-level diagrams. A common definition for such an asymmetry is $\afb\equiv\frac{N^{\Delta y>0}-N^{\Delta y<0}}{N^{\Delta y>0}+N^{\Delta y>0}},$ where $\Delta y\equiv y_t-y_{\bar t}$, $y_t$~($y_{\bar t}$) is the rapidity of the $t$~($\bar t$) quark, and $y=\frac12\ln\frac{E+p_z}{E-p_z}$. 

D0 measured \afb\ in the $\ttbar$ rest frame in lepton$+$jets final states on a dataset corresponding to 4.3~\fb\ using $\ttbar$ event candidates fully reconstructed with a kinematic fitter, and found $\afb=8\%\pm4\%$~\cite{bib:asymm_d0}. D0's result, shown in Fig.~\ref{fig:asymm}~(a), is about 2 standard deviations (SD) away from the {\sc mc@nlo}~\cite{bib:mcnlo} prediction of $1\pm2\%$. A similar measurement was performed by CDF in lepton$+$jets final states~\cite{bib:asymm_lj_cdf}, where additional tests for invariance under charge and parity conjugation are carried out, and the distribution in \afb, shown in Fig.~\ref{fig:asymm}~(b), is corrected back to parton level. After all corrections, CDF finds $\afb=16\%\pm7\%$, which is about 1.5~SD away from the {\sc mc@nlo} prediction of $6\%\pm1\%$. CDF also investigated the dependence of $\afb$ on the invariant mass of the $\ttbar$ system, $M_{\ttbar}$, which is compared to the prediction of {\sc mc@nlo} plus backgrounds in Fig.~\ref{fig:asymm}~(c). Motivated by the resolution in~$M_{\ttbar}$, CDF measures \afb\ in two bins of $M_{\ttbar}<450~\GeV$ and $M_{\ttbar}>450~\GeV$. %The result at parton level and after all corrections is shown in Fig.~\ref{fig:asymm}~(d). 
CDF finds that $\afb=48\%\pm11\%$ at parton level and after all corrections in the $M_{\ttbar}>450~\GeV$ bin is $>3$ SD away from the NLO SM prediction of~$\afb=9\%\pm1\%$. Another measurement, carried out by CDF in the dilepton channel using 5.1~\fb\ of data, yields $\afb=42\%\pm16\%$ parton level after all corrections~\cite{bib:asymm_ll_cdf}, which is about 2.5~SD away from the SM NLO prediction. The above results indicate tension between the measurement and the NLO SM prediction. Several mechanisms originating from new physics contributions have been suggested to explain this discrepancy, however, it has been pointed out that non-vanishing and acceptance-dependent contributions at higher orders in $\alpha_s$ within the SM could play an important role in understanding these findings.

\section{Conclusions}
Several measurements of key properties of the top quark were presented, most of which are in good agreement with the SM expectations. The forward-backward  asymmetry $\afb$ of $\ttbar$ production displays notable tension between the measuremnets and the SM NLO calculations. Both CDF and D0 expect to acquire about 12~\fb\ of data by the end of Run II of the Tevatron in September 2011, and we are looking forward to updates of the exciting measurements presented here with the full dataset.

%\begin{table}[t]
%\caption{Experimental Data bearing on $\Gamma(K \ra \pi \pi \gamma)$
%for the $\ko_S, \ko_L$ and $K^-$ mesons.\label{tab:exp}}
%\vspace{0.4cm}
%\begin{center}
%\begin{tabular}{|c|c|c|l|}
%\hline
%& & & \\
%&
%$\Gamma(\pi^- \pi^0)\; s^{-1}$ &
%$\Gamma(\pi^- \pi^0 \gamma)\; s^{-1}$ &
%\\ \hline
%\mco{2}{|c|}{Process for Decay} & & \\
%\cline{1-2}
%$K^-$ &
%$1.711 \times 10^7$ &
%\begin{minipage}{1in}
%$2.22 \times 10^4$ \\ (DE $ 1.46 \times 10^3)$
%\end{minipage} &
%\begin{minipage}{1.5in}
%No (IB)-E1 interference seen but data shows excess events relative to IB over
%$E^{\ast}_{\gamma} = 80$ to $100MeV$
%\end{minipage} \\
%& & &  \\ \hline
%\end{tabular}
%\end{center}
%\end{table}

\section*{Acknowledgments}
I would like to thank my collaborators from the CDF and D0 experiments for their help in preparing this article. I also thank the staffs at Fermilab and collaborating institutions, as well as the CDF and D0 funding agencies.


\section*{References}
\begin{thebibliography}{99}

\bibitem{bib:topdiscovery}
F. Abe \etal\ (CDF Coll.), Phys. Rev. Lett. {\bf 74}, 2626 (1995), 
S.~Abachi~\etal\ (D0 Coll.), Phys. Rev. Lett. {\bf 74}, 2632 (1995).

\bibitem{bib:singletop}
T. Aaltonen \etal\ (CDF Coll.), Phys. Rev. Lett. {\bf 103}, 092001 (2009),
V.~M.~Abazov~\etal\ (D0 Coll.), Phys. Rev. Lett. {\bf 103}, 092002 (2009).
%\bibitem{bib:topmassd0nature}V. Abazov \etal\ (D0 Collaboration),
%Nature \textbf{429}, 638 (2004).

\bibitem{bib:mass}
Z. Ye, these proceedings.

\bibitem{bib:toprescdf}
\verb|http://www-cdf.fnal.gov/physics/new/top/public_mass.html|
%\verbatim{}

\bibitem{bib:topresd0}
\verb|http://www-d0.fnal.gov/Run2Physics/WWW/results/top.htm|,
\verb|http://www-d0.fnal.gov/Run2Physics/WWW/documents/Run2Results.htm|.

\bibitem{bib:xsec}
S. Amerio, these proceedings.

\bibitem{bib:singletoptalk}
V. Bazterra, these proceedings.

\bibitem{bib:width_d0}
V. M. Abazov \etal\ (D0 Coll.), Phys. Rev. Lett. {\bf 106}, 022001 (2011).

\bibitem{bib:width_cdf}
T. Aaltonen \etal\ (CDF Coll.), Phys. Rev. Lett. {\bf 105}, 232003 (2010).

\bibitem{bib:spin_d0}
V. M. Abazov \etal\ (D0 Coll.), Fermilab-Pub-11/052-E, submitted to Phys.~Rev.~Lett.~B, arXiv:1103.1871 [hep-ex] (2011).

\bibitem{bib:spin_cdf}
T. Aaltonen \etal\ (CDF Coll.), CDF Conf. Note 10211 (2010).

%\bibitem{bib:spin_ll_cdf}

\bibitem{bib:whel_d0}
V. M. Abazov \etal\ (D0 Coll.), Phys. Rev. D {\bf 83}, 032009 (2011).

\bibitem{bib:whel_cdf}
T. Aaltonen \etal\ (CDF Coll.), CDF Conf. Note 10333 (2010). 

\bibitem{bib:colour}
V. M. Abazov \etal\ (D0 Coll.), Phys. Rev. D {\bf 83}, 092002 (2011).

%\bibitem{bib:charge}
%T. Aaltonen \etal\ (CDF Coll.), CDF Conf. Note 10460 (2011).

\bibitem{bib:asymm_d0}
V. M. Abazov \etal\ (D0 Coll.), D0 Conf. Note 6062 (2010).

\bibitem{bib:mcnlo}
S.~Frixione and B.R. Webber, JHEP {\bf 0206}, 029 (2002), S. Frixione, P. Nason and B.R. Webber, JHEP {\bf 0308} 007  (2003).

\bibitem{bib:asymm_lj_cdf}
T. Aaltonen \etal\ (CDF Coll.), Fermilab-Pub-10-525-E, submitted to PRD,\\ arXiv:1101.0034 [hep-ex] (2010).

\bibitem{bib:asymm_ll_cdf}
T. Aaltonen \etal\ (CDF Coll.), CDF Conf. Note 10436 (2011).

%\bibitem{bib:combi}
%The Tevatron Electroweak Working Group and CDF and D0 Collaborations, [arXiv:hep-ex/1007.3178] (2010).

\end{thebibliography}
\end{document}